\documentclass{aa}
\usepackage[dvips]{graphicx}

\def\teff{$T_{\rm eff}$}
\def\lgg{$\log\,{g}$}
\def\vt{$\xi_{\rm t}$}
\def\kms{\,km\,s$^{-1}$}

\def\vr{$V_{\rm r}$}

\def\i{\,{\sc i}} \def\ii{\,{\sc ii}} \def\iii{\,{\sc iii}} 

\def\EqW{$W_\lambda$}
\def\vsini{$v \sin i$} 

 \newcommand{\vwa}{{\sc vwa}}           

\begin{document}


\title{Abundance analysis of two late A-type stars HD 32115 and HD 37594.
	\thanks
		{Based on observations obtained at the 2-m telescope of Peak Terskol
		Observatory near Elbrus mountain, International Center of Astronomical
		and Medical-Ecological Researches, Russia }}

\author{I. F. Bikmaev\inst{1}, T. A. Ryabchikova\inst{2,3}, H. Bruntt\inst{4}, F. A. Musaev\inst{5,6,7},  
        L. I. Mashonkina\inst{1}, E. V. Belyakova\inst{1}, V. V. Shimansky\inst{1}, 
        P. S. Barklem\inst{8}, \and G. Galazutdinov\inst{5,6}}

\institute {Department of Astronomy, Kazan State University,
            Kremlevskaya 18, 420008 Kazan, Russia (Ilfan.Bikmaev@ksu.ru)
\and Institute of Astronomy, Russian Academy of Sciences,
     Pyatnitskaya 48, 109017 Moscow, Russia (ryabchik@inasan.rssi.ru)
\and Institute for Astronomy, University of Vienna, T\"urkenschanzstrasse 17,
             A-1180  Vienna, Austria (ryabchik@tycho.astro.univie.ac.at)
\and Institute for Physics and Astronomy, University of Aarhus, Bygn. 520, DK-8000 Aarhus C, Denmark (bruntt@ifa.au.dk)	     
\and Special Astrophysical Observatory RAS, 369167 Nizhnij Arkhyz, Karachai-Circassian Republic, Russia (faig@sao.ru)
\and SAO Branch of Isaac Newton Institute, Santiago, Chile
\and International Centre for Astronomical, Medical and Ecological Researches  (ICAMER), 
      National Academy of Sciences of Ukraine, 361605 Peak Terskol, Kabardino-Balkaria, Russia  
\and Department of Astronomy and Space Physics, Uppsala University, Box 515, 751-20 Uppsala, Sweden (barklem@astro.uu.se)}
\date{Received ; accepted}

\offprints{T. Ryabchikova}
\authorrunning{I. Bikmaev et al.}
\titlerunning{Abundance analysis of HD~32115 and HD~37594}

\abstract{
We have performed abundance analysis of two slowly rotating, late A-type stars, 
HD~32115 (HR~1613) and HD~37594 (HR~1940), based on obtained echelle spectra 
covering the spectral range 
4000-9850 \AA. These spectra allowed us to identify an extensive line list for 31 
chemical elements, the most complete to date for A-type stars. Two approaches to 
abundance analysis were used, namely a ``manual'' (interactive) and a semi-automatic 
procedure for comparison of synthetic and observed spectra and equivalent widths. 
For some elements non-LTE (NLTE) calculations were carried out and the corresponding 
corrections have been applied. The abundance pattern of HD~32115 was found to be very
close to the solar abundance pattern, and thus may be used as an abundance standard 
for chemical composition studies in middle and late A stars.  Further, its H$\alpha$ 
line profile shows no core-to-wing anomaly like that found for cool Ap stars and 
therefore also may be used as a standard in comparative studies of the atmospheric 
structures of cool, slowly rotating Ap stars.  HD~37594 shows a metal deficiency 
at the level of -0.3 dex for most elements and triangle-like cores of spectral lines. 
This star most probably belongs to the $\delta$ Sct group.
\keywords{stars: atmospheres -- stars: abundances -- stars: individual: HD\,32115, HD\,37594 }}

\maketitle

\section{Introduction}

Solar photospheric abundances are usually used as a reference for any abundance study. 
However, it has been shown that the mean abundances for some elements obtained for 
large homogeneous groups of stars can differ significantly from solar abundances. For 
example, Luck \& Lambert (1981, 1985) and later Takeda \& Takada-Hidai (1998) obtained 
an oxygen deficiency of 0.3 dex relative to the Sun for a large group of A to K 
supergiants. There are some indications that the sun may be slightly iron-rich 
(+0.09 dex on average) as compared with solar-like stars of the same age 
(Gustafsson 1998).  Extensive abundance studies of the superficially normal 
sharp-lined B- and F-type dwarfs were performed by S.\ Adelman and co-workers 
(see Adelman et al.\ 2000 and references therein). They found that many of the 
investigated stars show Am-type phenomena, and only 5 stars with effective temperatures 
in the range 6700--9000 K have abundances close to the solar photospheric values. Their 
study was performed mainly in the blue spectral region (up to 5000 \AA), and therefore 
abundances of C, N, O, Na, S, K were absent or rather questionable. Varenne \& Monier (1999) 
published abundances for 48 A- and F-type dwarfs in the Hyades open cluster. They 
have only two stars hotter than 6650 K with \vsini\ $\leq$ 25 \kms. Because of the 
limited spectral regions that were observed
abundances for only 11 elements have been derived.  
Half of these abundances were based on 1 or 2 spectral lines.  

Most sharp-lined stars with temperatures in the range 6700--8000 K belong to different 
groups of peculiar stars: Ap, roAp, Am, or $\lambda$ Boo groups. They are usually 
believed to have the same or nearly the same atmospheric structure as normal stars. 
However, Cowley et al.\ (2001) found a pronounced core-to-wing anomaly in the Balmer 
lines of some Ap stars. They could not compare the H$\alpha$ line profiles of these 
stars with normal solar abundance stars because of the lack of the reliable spectroscopic 
standards in this temperature region. We therefore decided to perform an accurate 
spectroscopic investigation of sharp-lined late A- to early F-type stars which are 
classified as normal MS stars, using observations of a wide spectral region 4000--9850 \AA.  
From tabulated rotational velocities (Abt \& Morrell 1995) we extracted 27 stars 
classified as normal A3--F5-type stars of luminosity classes III-V with \vsini$\leq$ 25 \kms. 
Among them only 13 stars have metallicities in the range $|[M]|\leq$0.15 on the basis of 
temperature-gravity-metallicity calibrations of the observed Str\"{o}mgren colours 
(a procedure which will be described below). We did not consider the remainder of the 
stars, the majority of which have low metallicities $[M]<$-0.15. In this paper we present 
the results of careful atmospheric parameter and abundance determinations for two stars: 
HD~32115 and HD~37594. These stars were classified as A8IV (HD~32115) and A8Vs 
(HD~37594) by Cowley et al.\ (1969).   

\section{Observations and data reduction}

High resolution spectra of the region 4000-9850 \AA\, were obtained for both 
stars with the coude-echelle spectrometer (Musaev~et~al.\ 1999) 
mounted on the 2\,m ``Zeiss'' telescope at Peak Terskol Observatory near Elbrus 
mountain in Russia. The best resolving power of the spectrograph in this 
operational mode is R=45\,000, but because of the rather poor seeing we used 
a wider slit which resulted in a reduced spectral resolution of 
R$\approx$30\,000.  A Wright Instruments front-side illuminated CCD of 1242 by 1152 
pixels (22 mkm) was used as a detector for simultaneous registration of the whole 
spectral range in a single exposure.  Two spectra were obtained for each star and 
they were coadded after cleaning for cosmic events during the reduction procedure. 
Blueward of 7100 \AA\, the consecutive orders overlap, while there are gaps of 5 to 
55 \AA\, between the orders for $\lambda\geq$7100 \AA. Typical S/N=150 was estimated 
for the centers of the orders close to the H$\alpha$ region, while S/N decreases 
to about 100 in the blue and infrared regions. Note that S/N also decreases from 
the center to both ends of the order. 
An extra spectrum of HD~32115 was obtained recently and was used for radial velocity
measurements, and for equivalent widths measurements of a few lines in the infrared
spectral region.
Table~1 contains the Heliocentric Julian Dates 
of the middle of the exposure for each pair of observations. The second column 
of Table~1 presents the results of the radial velocity measurements (see below) 
together with the data from the literature.

  \begin{table*}
	\caption{Observational log for HD\,32115 and HD\,37594.  
		 Radial velocity measurements based on $n$ lines with standard
		 deviations in units of the last digit (in parentheses) are given
		 in the second column.
		 }
							\label{RV}
	\begin{footnotesize}
	\begin{center}
	\begin{tabular}{|ccrc|}
	\noalign{\smallskip}
\hline	
HJD-2400000 & \vr,\kms & n & Reference\\						
\hline
  \multicolumn{4}{|c|}{HD 32115}\\
52333.2007 & 19.15(43) &  67 & this paper \\
51929.3035 & 42.08(45) &  71 & this paper \\
49405.5333 & 44.61(43) &     & Grenier et al.\ (1999) \\
42675.8313 & 43.7~(5~) &  14 & Nordstr\"{o}m \& Andersen (1985)\\
42669.8640 & 12.2~(4~) &  13 & Nordstr\"{o}m \& Andersen (1985)\\
42379.7278 & ~6.8~(6~) &  14 & Nordstr\"{o}m \& Andersen (1985)\\
  \multicolumn{4}{|c|}{HD 37594}\\
51930.2146 & 22.52(72) &  69 & this paper \\
49736.6142 & 22.04(39) &     & Grenier et al.\ (1999)          \\
42680.8220 & 22.4~(3~) &  15 & Nordstr\"{o}m \& Andersen (1985)\\
42675.8928 & 24.8~(6~) &  16 & Nordstr\"{o}m \& Andersen (1985)\\
42380.6909 & 20.2~(7~) &  16 & Nordstr\"{o}m \& Andersen (1985)\\
\hline
	\end{tabular}
	\end{center}
	\end{footnotesize}
	\end{table*}
	
Spectrum processing was realised with the help of a modified version of the PC-based 
DECH software (Galazutdinov 1992).  It includes background subtraction, echelle vectors 
extraction from the echelle-images, wavelength calibration, continuum rectification, 
line identification, equivalent widths and radial velocity measurements. We did not 
divide the stellar image by a flat-field as we found this procedure does not improve 
the initial S/N=100--150 for this particular CCD.

\section{Radial velocities}

Each spectral line chosen for measurement was approximated by a Gaussian. We measured 
332 and 268 lines in the spectra of HD~32115 and HD~37594 respectively. The full list 
of spectral lines in the 4000-9850 \AA\, spectral region for both stars was extracted 
from the {\sc vald} database (Kupka et al.\ 1999). For the final radial velocity we 
have chosen only Fe\i\, lines with the best laboratory wavelengths (Nave et al.\ 1994). 
Our \vr\, values and their standard deviations are given in the second column of 
Table~1 together with data collected from the literature. HD~37594 may be considered 
as a single star on the basis of six radial velocity measurements, while HD~32115 
shows clear \vr\, variations and may be a binary system. From the range of the radial 
velocities presented in Table 1 we may conclude that our first observations were undertaken 
near the maximum possible separation between the components. Careful synthesis of the 
whole spectrum of HD~32115 does not reveal any trace of the lines of the secondary star,
and therefore we conclude that it is most likely relatively faint and thus has no 
significant influence on the primary's spectrum.    

\section{Atmospheric parameters}

Stellar parameters derived with different methods are given in Table~2.  We first derived
effective temperatures, gravities, metallicities and absolute magnitudes of our stars
employing Str\"{o}mgren photometric indices and the calibration of Hauck \& Mermilliod
(1998). We used the {\sc templogg} procedure (Rogers 1995), which for A3--F0-type 
stars is based on the original calibrations by Moon \& Dworetsky (1985) with 
improvements by Napiwotski et al.\ (1993). For the metallicity calibration 
{\sc templogg} uses a relation between the Str\"{o}mgren metallicity index and the 
stellar metallicity derived by Smalley (1993).
The quoted formal errors in the stellar parameters from photometry account only for 
the reported accuracy of the photometric indices. One can see excellent agreement 
between absolute magnitudes obtained here and those obtained from Hipparcos 
parallaxes (ESA 1997). In this case the formal errors in the surface gravity include 
both the parallax uncertainty and an effective temperature uncertainty of $\pm$100 K. 
This provides strong support for the surface gravities used in our analysis. Bolometric 
corrections do not exceed 0.06 mag. On evolutionary tracks both stars lie close to the 
ZAMS and have masses slightly more than 1.5 $M_\odot$ (Pamyatnykh et al.\ 1998, Pamyatnykh 2000). 
The age of both stars is estimated at 600-700 Myr. 

To check the influence of the metal abundances on the derived atmospheric parameters 
we calculated a small grid of atmospheric models and corresponding colours with the 
abundances typical for our stars (see Sect. 5) and compared the observed and synthetic 
colours. Calculations were made with the {\sc atlas9} code (Kurucz 1993) which is 
modified by the inclusion of new line opacities in the form of specifically computed 
ODFs (Piskunov \& Kupka 2001) and the Canuto-Mazitelli convection treatment 
(Smalley \& Kupka 1997). We obtained the best fit to Str\"{o}mgren indices with 
\teff=7250 K and \teff=7170 K for HD~32115 and HD~37594, respectively. They are 
given in Table~2 as {\sc atlas9}-newODF results. The derived effective temperatures 
were also checked by fitting the H$\alpha$ line profile using these models.  
Special care was taken to achieve a correct continuum fit in the echelle orders 
containing the H$\alpha$ line. The typical length of the orders at 6400-6700 \AA\, 
is 90-100 \AA, and we have 2 overlapping orders containing the H$\alpha$ line. 
The estimated accuracy of our continuum level is $\pm$1-2$\%$.  The H$\alpha$ profiles 
of the models are computed as described in Barklem et al.\ (2000), namely, assuming 
LTE and using the {\sc synth} program (Piskunov 1992) with Stark broadening 
calculations from Stehl\'e \&\ Hutcheon (1999) and self-broadening from 
Barklem et al.\ (2000).  Radiative broadening and an estimate of the pressure 
broadening by helium are also included.  A comparison between the observed and 
calculated H$\alpha$ line profiles for HD~32115 is shown in Fig.1. 
Plots for HD~37594 show equally satisfactory fits. No atmospheric 
model can reasonably fit the core of the H$\alpha$ line in both stars, which is always broader in 
the observations.  Atmospheric parameters finally adopted for the abundance analysis 
are given in Table~2.

\begin{figure}[th]
\includegraphics[width=88mm]{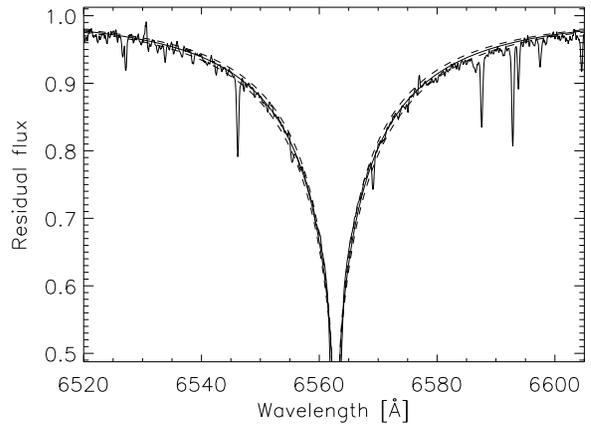}
\caption{A comparison between the observed H$\alpha$ line profile for HD~32115 
(thick full line) and those calculated for 3 atmospheric models: 7250\_newODF 
(adopted model -- thin full line), 7150\_newODF (upper dashed line) and 
7350\_newODF\label{fig1}.}
\end{figure}  

A further important parameter in the abundance analysis is microturbulent velocity. 
It was derived from the elements with numerous lines and the most accurate atomic data. 
Typically the \vt\ value as derived from lines of Cr, Mn, Fe, Ni, varies from 2.0 to 
2.6 \kms\, in HD~32115, and from 2.2 to 2.8 \kms\, in HD~37594. Ti lines give a 
systematically higher microturbulent velocity in both stars, up to 3.2 \kms.  
For abundance calculations we accepted \vt=2.3$\pm$0.3 \kms\, for HD~32115, and 
\vt=2.5$\pm$0.3 \kms\ for HD~37594.  These values for turbulent velocity fields 
in late A stars agree very well with theoretical predictions (Kupka \& Montgomery 2002).

  \begin{table*}
	\caption{Atmospheric parameters for HD\,32115 and 
		HD\,37594 with formal error
		 estimates in units of the last digit in parentheses.}
							\label{RV}
	\begin{footnotesize}
	\begin{center}
	\begin{tabular}{|lccrc|}
	\noalign{\smallskip}
\hline	
Method & \teff   & \lgg   & [M/H] & $M_{\rm v}$\\						
\hline
  \multicolumn{5}{|c|}{HD 32115}\\
Photometry         & 7300(40)~& 4.21(01) & ~0.03(04)  & 2.84(02) \\
Parallax           &          & 4.24(05) &            & 2.84(10) \\
H$\alpha$          & 7250(100)&          &            &          \\
{\sc atlas9}-newODF& 7250~~~~ & 4.30~~~~ &            &          \\
Adopted            & 7250~~~~ & 4.20~~~~ & ~0.00 ~~~~ &          \\
  \multicolumn{5}{|c|}{HD 37594}\\
Photometry         & 7170(50)~& 4.21(03) & -0.15(04)  & 2.98(07) \\
Parallax           &          & 4.22(05) &            & 2.90(07) \\
H$\alpha$          & 7100(100)&          &            &          \\
{\sc atlas9}-newODF& 7170~~~~ & 4.25~~~~ &            &          \\
Adopted            & 7170~~~~ & 4.20~~~~ & -0.30~~~~  &          \\
\hline
	\end{tabular}
	\end{center}
	\end{footnotesize}
	\end{table*}

\section{Abundance analysis}

To begin the abundance analysis synthetic spectra is computed for the whole observed 
region with the adopted atmospheric parameters and solar abundances. All lines that 
have depths greater than 0.5 \%\ of the continuum flux in the synthetic spectrum were 
extracted from the {\sc vald} database (Kupka et al.\ 1999). For a few elements it was 
necessary to make corrections to the atomic line data, and these corrections were made 
during the final steps of abundance analysis (see sections below for individual elements 
for details). The spectrum synthesis code {\sc synth} (Piskunov 1992) was used in all 
synthetic spectrum calculations. Using these calculations we estimated rotational 
velocities for both stars, which were found to be \vsini=9$\pm$2 \kms\, for HD~32115 
and  \vsini=17$\pm$2 \kms\, for HD~37594. 
While for the most part rotational plus instrumental broadening provides adequate fits 
to unblended spectral lines in HD~32115, this is not the case for HD~37594. Single 
spectral lines in this star often have V-shaped profiles rather than the usual U-shaped 
profiles produced by the rotational broadening. 
Similar line profiles are observed in 
pulsating stars (eg.\ the study of the $\delta$~Sct-type star FG Vir by Mittermayer 2001). 
This effect is marginal, and needs a further study with higher spectral resolution.

The final abundance analysis was done by three different methods: classical analysis 
using equivalent width data, spectral synthesis for strong and partially blended lines, 
and a semi-automatic procedure which uses both equivalent widths and spectrum synthesis 
techniques. The results of the first two methods were combined. We found no systematic 
difference between equivalent widths measured in the Solar Flux Atlas 
(Kurucz et al.\ 1984) and in a day-sky spectra obtained at the same dates and with the 
same spectrograph that was used for the taking the spectra of the program stars. The 
range of the errors in equivalent width measurements is 2--5 m\AA\, and depends on S/N 
ratio (position of the line on the echelle order), and blending effects. For most of 
the unblended lines this error does not exceed 2--3 m\AA. 
We also compared two independent sets of equivalent width measurements in HD~32115 
spectrum made by I.Bikmaev with the Gaussian approximation and by H. Bruntt with
the Voigt approximation. For 122 lines the mean difference between two sets is 
1.2 m\AA\ with a standard deviation 2.8 m\AA.

\subsection{Classical abundance analysis}

Comparing the synthetic and the observed spectra we selected spectral lines which are 
free of substantial blending effects and measured their equivalent widths. The final 
abundances were calculated with Kurucz's {\sc width9} code modified by V.\ Tsymbal 
for {\sc vald}-format input files. For a few elements, like Ba\ii\, which has mostly 
very strong lines, or Eu\ii\, and Nd\iii\, which have only 1--2 partially blended 
lines in the whole spectral region, abundances were derived with synthetic spectrum 
calculations. Final abundances for 31 chemical elements (41 ions) are collected in 
Table~3 (columns 2 and 7 for HD~32115 and HD~37594 respectively). The standard deviation 
is given in units of the last digit in parentheses. Solar abundances in 
$\log(N/N_{tot})$ = $\log(N/N_H)$ - 0.04 (correction due to He) are taken from Grevesse \& Sauval (1998) 
except C, N, O, Mg, Si, and Ba. For these elements, except Ba, the most recent data by 
Holweger (2001) are given. We adopted the solar Ba abundance derived by Mashonkina \& Gehren (2000).  
NLTE corrections were applied for the species marked by asterisks (see details below). 
Columns 6 and 11 contain the stellar abundance relative to the sun.  We also calculated 
line-by-line differences between stellar and solar abundances, which show a close 
agreement with the values from columns 6 and 11.  All individual line abundances are presented in 
Table 4 which is available in electronic form only.  
This table may be retrieved from {\it http://www.inasan.rssi.ru/\verb"~"ryabchik}.
NLTE corrections are also included in Table 4. These are 
probably the most complete abundance data for single A-type stars to date.

\subsection{Semi-automatic abundance analysis}

We have written software called \vwa\ for semi-automatic abundance analysis which is 
described in Bruntt et al.\ (\cite{hd49434}). The code runs though the list of atomic 
line data extracted from {\sc vald} (Kupka et al.\ 1999) for the target star, and 
selects lines that suffer from the least amount of blending. For a region around each 
selected line the synthetic spectrum is calculated for different abundances of the 
element forming the line until the equivalent widths of the observed and
synthetic 
spectrum match. The computation of the synthetic spectrum is done with {\sc synth} 
version 2.5 (Valenti \& Piskunov 1996). The code needs the line data from {\sc vald} 
and the adopted atmosphere model as input. When \vwa\ has fitted the selected lines 
they are inspected visually -- and the fit is either accepted, rejected or must be 
improved manually. 

While classical abundance analysis can be performed relatively fast for a star like 
HD~32115 which has low \vsini\ the advantage of using \vwa\ for abundance analysis 
is demonstated for HD~37594. Since this star has a moderate value of \vsini\ one needs 
to select lines with some amount of blending to be able to derive abundances of as many 
elements as possible. We first used \vwa\ where we only selected unblended Fe lines. 
We then adjusted the microturbulence and recalculated abundances with \vwa\ until the 
Fe\i\ abundance and equivalent width of the lines did not correlate. We thus found 
$\xi_t = 2.4\pm0.3$ \kms. Since the abundance of Fe was now considered to be accurate, 
we allowed \vwa\ to select lines of other elements, which were mild blends with Fe. 
By this method we were able to use lines that could not be used in the classical 
abundance analysis which is strictly limited to unblended lines.

When comparing the results of the abundance analysis of the two methods in Table 3 we 
see that far fewer lines are chosen by \vwa\, but the derived abundances all agree within 
the error estimates. Hence, \vwa\ is usable for carrying out fast and reliable 
abundance analysis.

\subsection{Errors in abundance determinations}

Our derived abundances contain errors from a number of sources of uncertainty, 
including the adopted atomic data, most importantly the oscillator strengths, the 
equivalent width measurements, the adopted atmospheric parameters and model, and the 
observed spectra. For some sources we can directly estimate the associated error.  
The estimated uncertainty in equivalent width measurements is $\pm$2 m\AA, which 
results in 0.20--0.09 dex error in the abundance for lines with equivalent widths 5 to 
10 m\AA, 0.05 dex for \EqW=20 m\AA, and less than 0.03 dex for \EqW$\geq$40 m\AA.  
An error of 100 K in the effective temperature determination results in a systematic 
error of 0.05 dex on the the abundance of the most temperature-sensitive atomic species, 
e.g. Ti\i, V\i. For cases where a suitably large number of lines are used, the standard 
deviations quoted in Table~3 estimate the total random error associated with all sources 
of uncertainty.  However, for species where only a small number of lines are available 
this cannot be expected to give any indication of this 
error.  One can see from Table~3 that the standard deviations for most elements with a 
sufficient number of spectral lines do not exceed 0.15 dex.  Considering this as 
indicative for all elements, and taking into account also possible systematic errors 
we estimate the total error at about 0.2 dex for the species with 1--3 spectral lines.

Formally, a gravity decrease by 0.2 dex may provide ionization balance for all iron 
peak elements, since abundances obtained from the ionized lines will be decreased by 
0.06-0.09 dex, while abundances from the neutral lines would not be changed. CNO 
abundances would be reduced by 0.05 dex. However, we stress that the effective gravity 
for both stars is accurately defined by parallax measurements.  

Since most solar photospheric abundances are derived with the semi-empirical model of 
Holweger \& M\"uller (1974), it is important to check that our use of theoretical 
Kurucz models does not itself introduce any significant differences relative to the 
solar abundances. To test this we compared solar Fe abundances derived from Fe\ii\ 
lines in the solar spectrum employing the Holweger-M\"{u}ller solar atmospheric model 
(Holweger \& M\"{u}ller 1974) with those found using the solar model computed with 
{\sc atlas9} and the same convection treatment as for our stars (see Gardiner et al.\ 1999). 
The abundances were calculated from the best 13 Fe\ii\, lines (Schnabel, Kock \& Holweger 1999) 
with oscillator strengths from their paper and from the {\sc vald} database. 
The results are:
$\log N_{\rm Fe}=$7.42$\pm$0.09 (SKH $\log gf$ values, HM model, weighted mean)), 
$\log N_{\rm Fe}=$7.46$\pm$0.11 (SKH $\log gf$ values, our model, unweighted mean),
$\log N_{\rm Fe}=$7.48$\pm$0.10 ({\sc vald} $\log gf$ values, our model, unweighted mean). 
With our solar model the line-by-line difference (HD~32115$-$Sun) for 31 common Fe\ii\, 
lines is -0.02$\pm$0.10 and agrees within the error limits with the [Fe/H] value 
presented in Table 3.

 \begin{table*}
	\caption{Abundances in HD\,32115 and HD\,37594
		 based on $n$ measured lines with the standard deviation
		 in units of the last two digits in parentheses. NLTE
		 corrections are applied for the species marked by asterisks.
		 The letters 'H' and 'MG' in the last column mean that solar abundances are
		 taken from Holweger (2001) and Mashonkina \& Gehren (2000); otherwise standard solar composition 
		 as reported by Grevesse \& Sauval (1998) is used.}
							\label{Abund}
	\begin{footnotesize}
	\begin{center}
	\begin{tabular}{|l|cc|cc|c|cc|cc|c|c|}
	\noalign{\smallskip}
\hline
Ion &\multicolumn{5}{c|}{HD 32115} &\multicolumn{5}{c|}{HD 37954}& Sun\\   
\hline
    &\multicolumn{2}{c|}{Classical analysis} &\multicolumn{2}{c|}{Semi-automatic }& &\multicolumn{2}{c|}{Classical analysis}&\multicolumn{2}{c|}{Semi-automatic }& & \\
    &$\log(N/N_{tot})$&$n$&$\log(N/N_{tot})$&$n$&[$N/N_H$]&$\log(N/N_{tot})$&$n$&$\log(N/N_{tot})$&$n$&[$N/N_H$]&$\log(N/N_{tot})$\\
\hline                                                                                                               
 ~C I   & ~-3.62(17) &~35 & -3.52(24)  &~~5 & -0.17 & ~-3.73(15) &~27 & -3.74(05) &~~4 &-0.28 & ~-3.45H~\\
 ~N I   & ~-4.19(08) &~11 & -4.22(04)  &~~2 & -0.08 & ~-4.27(12) &~~4 & 	  &    &-0.16 & ~-4.11H~\\
 ~O I   & ~-3.23(10) &~12 & -3.25(04)  &~~2 & ~0.07 & ~-3.36(12) &~~5 & -3.27(12) &~~5 &-0.06 & ~-3.30H~\\
 Na I*  & ~-5.86(12) &~~8 &	       &    & -0.15 & ~-6.05(21) &~~8 & 	  &    &-0.34 & ~-5.71~~\\
 Mg I   & ~-4.59(12) &~~9 & -4.33(05)  &~~4 & -0.09 & ~-4.89(10) &~~8 & -4.68(03) &~~4 &-0.39 & ~-4.50H~\\
 Mg II  & ~-4.43(11) &~~4 & -4.46~~~~  &~~1 & ~0.07 & ~-4.79(19) &~~3 & -4.65~~~  &~~1 &-0.29 & ~-4.50H~\\
 Al I   & ~-5.69(15) &~~4 &	       &    & -0.12 & ~-6.15(06) &~~2 & 	  &    &-0.58 & ~-5.57~~\\
 Si I   & ~-4.53(13) &~39 & -4.49(12)  &~11 & -0.03 & ~-4.75(08) &~26 & -4.76(13) &~11 &-0.25 & ~-4.50H~\\
 Si II  & ~-4.47(12) &~~6 & -4.53~~~~  &~~1 & ~0.03 & ~-4.59(04) &~~3 & -4.72(10) &~~2 &-0.09 & ~-4.50H~\\
 ~P I   & ~-6.42(13) &~~2 &            &    & ~0.17 &            &    &           &    &      & ~-6.59~~\\
 ~S I   & ~-4.81(15) &~19 & -4.92(09)  &~~5 & -0.10 & ~-5.07(10) &~~8 & -5.06(01) &~~2 &-0.36 & ~-4.71~~\\
 Cl I   & ~-6.76~~~~ &~~1 &	       &    & -0.22 &   	 &    & 	  &    &      & ~-6.54~~\\
 ~K I*  & ~-7.05~~~~ &~~1 &	       &    & -0.13 & ~-7.32~~~~ &~~1 & 	  &    &-0.40 & ~-6.92~~\\
 Ca I   & ~-5.63(15) &~21 & -5.60(13)  &~13 & ~0.05 & ~-5.89(22) &~26 & -5.99(27) &~16 &-0.21 & ~-5.68~~\\
 Ca II* & ~-5.69(11) &~~7 & -5.58~~~~  &~~1 & -0.01 & ~-5.89(04) &~~4 & -5.88~~~~ &~~1 &-0.21 & ~-5.68~~\\
 Sc II  & ~-8.89(10) &~11 & -8.96(20)  &~~4 & -0.02 & ~-9.04(15) &~~6 & -9.13~~~~ &~~3 &-0.17 & ~-8.87~~\\
 Ti I   & ~-7.20(14) &~21 & -7.18(04)  &~~4 & -0.18 & ~-7.43(14) &~~3 & -7.42(03) &~~2 &-0.41 & ~-7.02~~\\
 Ti II  & ~-7.03(15) &~27 & -6.92(20)  &~13 & -0.01 & ~-7.23(18) &~19 & -7.38(12) &~~8 &-0.21 & ~-7.02~~\\
 ~V I   & ~-8.24(25) &~~3 &	       &    & -0.20 &   	 &    & 	  &    &      & ~-8.04~~\\
 ~V II  & ~-8.17(08) &~~5 & -7.95~~~~  &~~1 & -0.13 & ~-8.13(25) &~~2 & 	  &    &-0.09 & ~-8.04~~\\
 Cr I   & ~-6.40(12) &~39 & -6.40(13)  &~~7 & -0.03 & ~-6.71(10) &~11 & -6.75(06) &~~5 &-0.34 & ~-6.37~~\\
 Cr II  & ~-6.27(09) &~20 & -6.22(12)  &~~7 & ~0.10 & ~-6.63(09) &~17 & -6.55(06) &~~6 &-0.26 & ~-6.37~~\\
 Mn I   & ~-6.87(17) &~20 & -6.76(37)  &~~6 & -0.22 & ~-7.14(16) &~~9 & -7.04(13) &~~4 &-0.49 & ~-6.65~~\\
 Mn II  & ~-6.72(11) &~~2 &	       &    & -0.07 &   	 &    & 	  &    &      & ~-6.65~~\\
 Fe I   & ~-4.60(12) &235 & -4.59(11)  &109 & -0.06 & ~-4.88(13) &183 & -4.89(12) &~38 &-0.34 & ~-4.54~~\\
 Fe II  & ~-4.48(14) &~60 & -4.54(11)  &~23 & ~0.06 & ~-4.87(11) &~25 & -4.76(11) &~~4 &-0.33 & ~-4.54~~\\
 Co I   & ~-7.16(10) &~~5 & -7.16(00)  &~~2 & -0.04 & ~-7.43(03) &~~3 & 	  &    &-0.31 & ~-7.12~~\\
 Ni I   & ~-5.91(10) &~59 & -5.94(13)  &~17 & -0.12 & ~-6.24(16) &~27 & -6.18(03) &~~4 &-0.45 & ~-5.79~~\\
 Ni II  & ~-5.82~~~~ &~~1 &	       &    & -0.03 &   	 &    & 	  &    &      & ~-5.79~~\\
 Cu I   & ~-7.82(26) &~~5 &	       &    & ~0.01 & ~-8.32(06) &~~2 & 	  &    &-0.49 & ~-7.83~~\\
 Zn I   & ~-7.77(16) &~~3 &	       &    & -0.33 & ~-8.07(05) &~~3 & 	  &    &-0.63 & ~-7.44~~\\
 Sr II* & ~-9.09(22) &~~3 &	       &    & -0.02 & ~-9.19(07) &~~3 & 	  &    &-0.12 & ~-9.07~~\\
 ~Y II  & ~-9.68(15) &~~8 & -9.74(08)  &~~3 & ~0.12 & ~-9.94(08) &~~8 & -9.95(07) &~~3 &-0.14 & ~-9.80~~\\
 Zr II  & ~-9.38(16) &~~7 &	       &    & ~0.06 & ~-9.42(20) &~~5 & 	  &    &~0.02 & ~-9.44~~\\
 Ba II* & ~-9.64(12) &~~6 & -9.57(05)  &~~3 & ~0.19 & ~-9.87(28) &~~4 & -9.70(19) &~~4 &-0.04 & ~-9.83MG\\
 La II  & -10.82(08) &~~4 &	       &    & ~0.05 & -11.15(13) &~~3 & 	  &    &-0.28 & -10.87~~\\
 Ce II  & -10.41(19) &~~7 &	       &    & ~0.05 & -10.57(16) &~~6 & 	  &    &-0.11 & -10.46~~\\
 Nd II  & -10.49(25) &~11 &	       &    & ~0.05 & -10.63(01) &~~2 & 	  &    &-0.09 & -10.54~~\\
 Nd III & -10.40(07) &~~2 &	       &    & ~0.14 &   	 &    & 	  &    &      & -10.54~~\\
 Sm II  & -11.06~~~~ &~~1 &	       &    & -0.03 & -11.38~~~~ &~~1 & 	  &    &-0.35 & -11.03~~\\
 Eu II  & -11.63~~~~ &~~1 &	       &    & -0.10 & -11.80~~~~ &~~1 & 	  &    &-0.27 & -11.53~~\\
\hline       
	\end{tabular}
	\end{center}
	\end{footnotesize}
	\end{table*}

\subsection{CNO elements}

Oscillator strengths for C and N were taken from Hibbert et al.\ (1993) and Hibbert et al.\ (1991) 
respectively. The same data were used in reevaluation of the solar abundances 
(Holweger 2001). For oxygen, we employed the oscillator strengths supplied by 
{\sc vald} which originate from the NIST compilation (Wiese et al.\ 1996).  We did 
not use the IR O\i\, triplet because of the possibly strong NLTE effects. Expected 
NLTE corrections are -0.1 dex for C (see Sturenburg, 1993, Rentzsch-Holm 1996, Paunzen et al.\ 1999) 
and -0.01 -- -0.03 dex for O (Takeda 1997). Both stars show similar small deficiencies 
of C and N.

\subsection{Na to K}

These elements are known to be subject to NLTE effects. We also note that the oscillator 
strengths for Si\i\, lines in {\sc vald} are not accurate, many of them coming from 
Kurucz \& Peytremann (1975). Therefore we recalculated oscillator strengths for most 
of the Si\i\, lines by fitting them to the Solar Flux Atlas (Kurucz et al.\ 1984) 
adopting the solar abundance in Table~3. NLTE corrections were calculated for Na\i, 
and K\i\, following atomic models and procedures developed by Mashonkina et al.\ (2000) 
for Na, and Ivanova \& Shimansky (2000) for K. Oscillator strength data for Al\i\ 
were taken from Baum\"{u}ller \& Gehren (1996). According to this paper NLTE corrections 
for the Al\i\ lines used in our abundance analysis are negligible. For Mg\i\ lines 
$\lambda\lambda$ 4703, 4730, 5528, 5711 oscillator strengths were taken from Jonsson et al.\ (1984) 
and from Froese Fischer (1975). In both stars the Mg abundance was derived by the 
spectral synthesis technique with the Stark and Van der Waals damping constants giving 
the best fit for the same Mg\i\ lines in the spectra of Procyon and the Sun 
(Fuhrmann et al.\ 1997, Zhao et al.\ 1998). Individual LTE and NLTE abundances are 
given in Table~4. The largest NLTE corrections were found for the K\i\, resonance 
line $\lambda$~7698.97. They are -0.61 (HD~32115) and -0.41 (HD~37594). Unfortunately, 
another K\i\, resonance line at $\lambda$~7664.91 falls in one of the gaps in our 
echelle spectra. We could measure the strongest Cl\i\, line $\lambda$~8375.95 in the 
spectrum of HD~32115. The solar photospheric Cl abundance is very uncertain, 
$\log(Cl/N_{tot})$=-6.54$\pm$0.3. The Cl abundance in HD~32115 was obtained from just 
one weak line but agrees perfectly with the Cl abundance found in meteorites 
(Grevesse \& Sauval 1998). Two strongest infrared P\i\ lines $\lambda\lambda$~9593.50,
9796.83 measured in the latest spectrum of HD~32115 were used to derive phosphorus abundance.  

\subsection{Ca to Ni}

The iron-group elements usually represent the metallicity of a star.  With a few 
exceptions we were able to obtain abundances from the lines of neutral and singly-ionized 
species for all elements of this group, thus checking the ionization balance in the 
atmospheres. The abundances obtained by the neutral and ionized Fe-group elements differ 
by $\approx$0.13 dex on average. In both our stars the Fe-group elements in the ionized 
state are dominating and therefore their lines are less sensitive to temperature 
inaccuracy and NLTE effects and hence should provide better abundance estimates. 
Among iron-group elements NLTE calculations were carried out for Ca (see below) and Fe 
(Rentzsch-Holm 1996). An expected NLTE correction to the abundance obtained from 
Fe\i\ lines is $\approx$+0.1 dex for \teff$\approx$7300 K. Applying this correction 
as a first approximation to the abundances obtained from the lines of neutral elements 
a reasonable ionization balance is obtained. 

Five Fe\ii\ lines correspondning to 4d-4f transitions with an upper energy level at 
around 10.5 eV were measured in the HD~32115 spectrum.  For these lines only theoretical 
oscillator strengths are available. Both the $\log gf$ values from the Kurucz line 
list (included in {\sc vald}) (Kurucz 1993) and the more accurate sets of calculated 
$\log gf$-values from Raassen \& Uylings (1998) agree to within 0.07 dex. Abundances 
obtained from these lines do not deviate by more than 0.1-0.2 dex from the mean Fe 
abundance and thus supports the oscillator strengths for these high excitation Fe\ii\ 
lines to this level of accuracy.

Fe abundance analysis in HD~32115 is based on about 300 lines well spread through the 
whole spectral region of 4100--9200 \AA. We did not find any clear dependence of the 
individual abundances on the wavelength (Fig.2) which might be expected in the primary 
star if the contribution from the secondary is not negligible. 

For the two strongest Ca\ii\ lines $\lambda\lambda$ 8912, 8927 we increased $\log gf$ 
by 0.3 dex to fit line profiles in the solar spectrum. Stark damping constants for 
these lines, which are important, are taken from Dimitrijevi\'{c} \& Sahal-Br\'{e}chot (1993). 
We also carried out NLTE calculations for Ca\ii\ lines using a model of the atom 
developed by Ivanova et al.\ (2002). NLTE corrections are always negative and of 
the order of -0.15 dex for $\lambda\lambda$ 8248, 8912, 8927 lines in both stars, 
while they are negligible for weak Ca\ii\ lines. 

\begin{figure}[th]
\includegraphics[width=88mm]{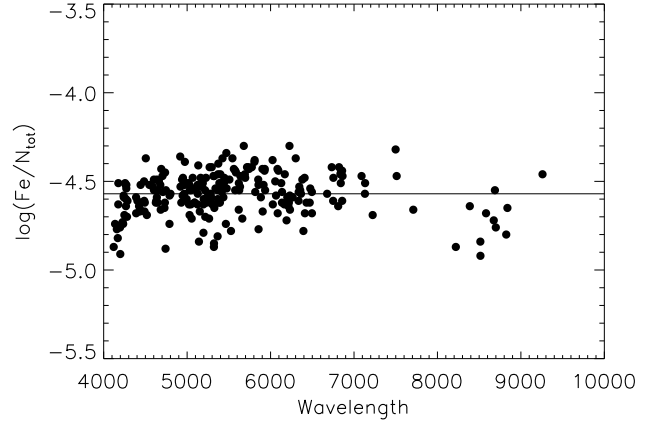}
\caption{The dependence of the individual abundances obtained from Fe\i\ lines in HD~32115 on wavelength.
}\label{fig2}
\end{figure}  
    
\subsection{Cu and Zn}

Abundances of these elements were obtained by spectral synthesis taking into account 
hyperfine splitting (hfs). Hfs constants were extracted from Biehl
(1976). Solar synthetic 
spectrum calculations showed that for Zn\i\ lines we may neglect the hfs effect, 
while it is strong for Cu\i\ lines in the solar spectrum. In both our stars Cu\i\ 
lines are weak, and the hfs effect for Cu\i\ and Zn\i\ lines is not significant. 
   
Zn shows the largest deviation from the solar abundance in HD~32115, and the largest 
deficiency in HD~37594.  To our knowledge, there are no NLTE calculations for Cu\i\ 
or Zn\i. With ionization energies $\chi$ = 7.72 eV for Cu\i\ and 9.39 eV for Zn\i\ 
these atoms are minor species similar to Mg\i, Al\i, Fe\i\ in the stellar atmospheres 
typical for investigated stars. Based on our experience in NLTE analyses and taking 
into account NLTE calculations for atoms with similar atomic parameters and term 
structures (Sturenburg 1993, Rentzsch-Holm 1996) Cu\i\ and Zn\i\ are expected to be 
over-ionized. In this case low atomic levels are underpopulated compared 
with LTE, and the lower the excitation energy is the greater the departures from LTE 
are. Most probably, NLTE effects will weaken the investigated Cu\i\ and Zn\i\ lines 
and NLTE abundance corrections are expected to be positive. Usually NLTE corrections 
are increasing with the line intensity. The observed equivalent widths of Zn\i\ lines 
are 2-3 times larger than for Cu\i\ lines, therefore we may also expect larger positive
NLTE corrections for Zn.

\subsection{Sr, Y, Zr, Ba}

NLTE corrections for Sr\ii\ lines were calculated using the model atom developed by 
Belyakova et al.\ (1997, 1999). They do not exceed 0.1 dex. Hfs effects may be 
important for Y\ii\ lines, but there are no data available. Zr abundances are rather 
uncertain, because reliable oscillator strengths are only known for a small number of 
spectral lines. We could not use the solar spectrum fitting for Zr\ii\ lines measured 
in our stars due to strong molecular contributions in the corresponding spectral 
regions.  

NLTE corrections to the Ba abundance were calculated according to Mashonkina et al.\ (1999). 
They are large, up to 0.18 dex, for the two strongest Ba\ii\ lines $\lambda\lambda$~6141, 6496. 
It should be mentioned that the solar Ba abundance used by us was derived by Mashonkina \& Gehren (2000) 
from the same lines as in our analysis, and it is closer to abundance of Ba found in 
meteorites than any previously published solar Ba abundance values (see Grevesse \& Sauval 1998). 
Stark damping constants for Ba\ii\ lines were taken from Dimitrijevi\'{c} \& Sahal-Br\'{e}chot (1996), 
while for resonance Sr\ii\ lines the corresponding values were taken from the NIST 
compilation (Konjevi\'{c} et al.\ (1984)).

\subsection{La to Eu}

New experimental oscillator strengths and hfs constants were used for La\ii\ 
(Lawler et al.\ 2001a), Ce\ii\ (Palmeri et al.\ 2000, Zhang et al.\ 2001), and Eu\ii\ lines 
(Lawler et al.\ 2001b).  Two weak lines of Nd\iii\ $\lambda\lambda$ 5102, 5295, which are the 
strongest lines in the Nd\iii\ spectrum, were identified in HD~32115. They are 
partially blended so abundance estimates were made by spectral synthesis. Oscillator 
strengths and other atomic parameters in {\sc vald} are taken from Cowley \& Bord (1998) 
and from Bord (2000). Both lines provide Nd abundances close to that obtained from 
the analysis of Nd\ii\ lines, thus giving strong support for the Nd\iii\ oscillator 
strength calculations.

\section{Discussion}

The abundances found for HD~32115 and HD~37594 relative to the sun are shown in Fig.3 
by filled and open circles respectively.  Within the typical error limits of $\pm$0.15 
dex HD~32115 is a solar abundance star (a mean metallicity $[M] = -0.04\pm0.11$). 
Therefore this star may be used as a chemical standard in studies of cool peculiar stars. 
We also get a standard for further investigation of the hydrogen wing-to-core anomaly 
found in cool Ap stars by Cowley et al.\ (2001). Fig.4 shows a comparison between 
H$\alpha$ line profiles in HD~32115 and in one of the pulsating Ap (roAp) stars, 
HD~24712, with the same effective temperature. It is evident from this comparison that 
the Ap-star anomaly occurs in that part of the hydrogen line core which is not reproduced 
by our current models. 

\begin{figure}[th]
\includegraphics[width=88mm]{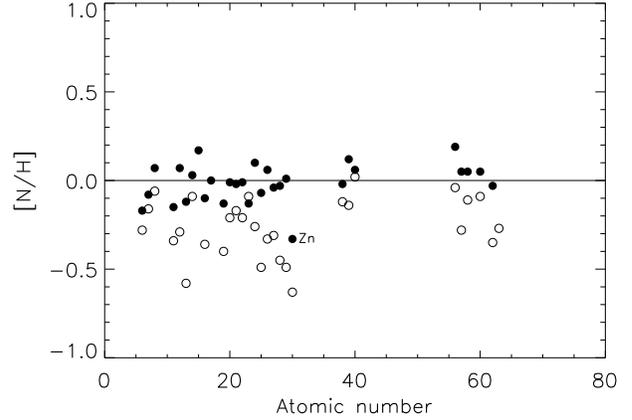}
\caption{The observed relative abundances in HD~32115 (filled circles) and in HD~37594 (open circles).}
\label{fig3}
\end{figure}  

\begin{figure}[th]
\includegraphics[width=88mm]{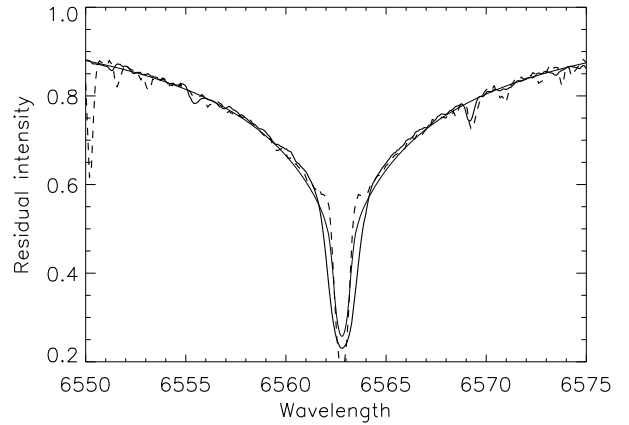}
\caption{A comparison between H$\alpha$ line profiles observed in HD~32115 (thick full line) and in
roAp star HD24712 (dashed line), and calculated for the adopted 7250-ODF model (thin full line).}
\label{fig4}
\end{figure}

HD~37594 is slightly metal-deficient. Its mean metallicity $[M]=-0.26\pm0.16$ is close 
to the value $[M]=-0.15\pm0.04$ obtained from photometric calibrations.  This error 
in the metallicity obtained from photometric calibrations is based only on the errors 
in the photometric indices. No systematic errors of the calibration itself are included, 
which may be up to 0.1.  
Further support for the high accuracy of the metallicities derived from the 
Str\"{o}mgren photometry for normal stars is provided by the results of abundance 
analysis of 28 And = HD~2628 (Adelman et al.\ 2000) and $\sigma$ Boo~A=HD~128167 
(Adelman et al.\ 1997). [M]=-0.16$\pm$0.29 and -0.29$\pm$0.37 were derived for these 
stars from abundances versus -0.13 and -0.35 from photometric calibration. The 
abundance pattern in HD~37594 is similar to that in 28 And (\teff=7250 K, \lgg=4.2, 
\vt=2.3 \kms, \vsini=9 \kms -- see Adelman et al.\ 2000) which belongs to the $\delta$ Sct group. 
The smaller scatter derived in the present paper is explained by more accurate abundance 
determinations with NLTE and hfs effects taken into account as well as more careful 
choice of the unblended lines which is possible in the red spectral region. 

Abundances obtained for HD~32115 and HD~37594 do not follow the predicted abundance 
pattern for $Z\leq$28 from consistent stellar evolution models calculated with 
radiative forces, opacities and diffusion (Turcotte et al.\ 1998). Models for 
1.5 $M/M_{\sun}$ predict that elements with 5~$<$~$Z$~$<$~20 will be underabundant 
by 0.5 dex relative to Fe, and that iron-peak elements will be generally slightly 
overabundant even at early evolutionary phases on the MS. This was not observed in 
either of the stars analysed here.

To confirm the derived abundances we have used the software 
\vwa\ (Bruntt et al.\ 2002) which was developed to
be able to make fast semi-automatic abundance analysis. 
We find the same abundances as the more careful classical approach. 
At present \vwa\ is being applied to the study 
of the primary target candidate stars for the asteroseismology missions
COROT and  R\o mer (Bruntt et al.\ 2002). Several of these stars have
moderate or high \vsini\, and the amount of work needed to carry out the
analysis using \vwa\ is reduced substantially.

\begin{acknowledgements}
We thank G.\ Wade who provided us with the spectrum of HD~24712 in H$\alpha$ region, 
A.\ Bondar for the help during the observations, N. Sakhibullin for useful discussions, 
and F.\ Kupka for his help in new model atmosphere calculations. T.\ R.\ thanks the 
Fonds zur F\"{o}rderung der wissenschaftlichen Forschung (project $P-14984$), 
the Jubil\"aumsfonds der \"Osterreichischen Nationalbank (project 7650), 
the Russian National Program ``Astronomy'', and RFBR (grant 00-15-96722) 
for financial support. IFB, LIM. EVB, VVS acknowledge the Russian National Program 
``Astronomy'', and  RFBR (grant  02-02-17174). FAM and GG thank the Russian National 
Program ``Astronomy'', and RFBR (grant 02-02-17423) for partial funding.
The authors thank a referee, C.R. Cowley, for his helpful comments. 
\end{acknowledgements}

\end{document}